\documentclass[aip,apl,reprint,amsmath,amssymb,floatfix]{revtex4-1}
\usepackage{epsf}
\usepackage{graphicx}
\usepackage{amsfonts}
\usepackage{dcolumn}%
\usepackage{color,ulem}
\usepackage{bm}

\begin{document}


\title{Towards precise measurement of oscillatory domain wall by ferromagnetic Josephson junction} 



\author{Shin'ichi Hikino}
\thanks{
JST, CREST, Sanbancho, Tokyo 102-0075, Japan}
\affiliation{
Computational Condensed Matter Physics Laboratory, RIKEN Advanced Science Institute (ASI), Wako, Saitama 351-0198, Japan}

\author{Michiyasu Mori}
\thanks{
JST, CREST, Sanbancho, Tokyo 102-0075, Japan}
\affiliation{
Advanced Science Research Center, Japan Atomic Energy Agency, Tokai, Ibaraki 319-1195, Japan}

\author{Wataru Koshibae}
\thanks{
JST, CREST, Sanbancho, Tokyo 102-0075, Japan}
\affiliation{
Cross-Correlated Materials Research Group (CMRG), RIKEN Advanced Science Institute (ASI), Wako, Saitama 351-0198, Japan}

\author{Sadamichi Maekawa}
\thanks{
JST, CREST, Sanbancho, Tokyo 102-0075, Japan}
\affiliation{
Advanced Science Research Center, Japan Atomic Energy Agency, Tokai 319-1195, Japan}

\date{\today}

\begin{abstract}
We theoretically propose a principle for precise measurement of oscillatory domain wall (DW) by a ferromagnetic Josephson junction,
which is composed of a ferromagnetic wire with DW and two superconducting electrodes. 
The current-voltage 
curve exhibits stepwise structures, only when DW oscillates in the ferromagnetic wire. 
The voltage step appears at $V=n(\hbar/2e)\omega_{\rm DW}$ with the fundamental constant $\hbar/e$, integer number $n$, and the DW frequency $\omega_{\rm DW}$.  
Since $V$ can be determined in the order of 10$^{-9}$ accuracy, the oscillatory DW will be measured 
more precisely than present status by conventional method. 
\end{abstract}


\maketitle 
Nano-scale magnetic materials for spintronics devices are extensively studied due to many advantages such as enhanced 
operation speed, 
low power consumption, and high integration of memory cell.\cite{zutic,maekawa} 
Non-volatile memory using a magnetic domain wall (DW) is one example of such devices,\cite{allwood, parkin} 
and many studies are devoted to control DW.\cite{yamanouchi,yamaguchi,saitoh,hayashi,boone,thiaville,he07,bisig,ieda}
Among those studies, the oscillatory DW is experimentally observed\cite{saitoh,hayashi,boone} 
and is examined toward 
applications, e.g. high sensitive magnetic sensor, nano-scale telecommunication, 
rf-assisted writing of magnetic bit, and microwave generator.\cite{kiselev,thirion,ono08,wang10}
Once such devices are realized in a microscopic circuit, one needs to measure the DW frequency more precisely. 

One example of precise measurement is the Josephson effect.\cite{josephson} 
Under irradiation of microwave to the Josephson junction,  
the current-voltage ($I$-$V$) curve shows step structures at $V=n(\hbar/2e)\nu$ with microwave frequency $\nu$, integer $n$, 
and the ratio of the Plank constant and the elementary charge $\hbar/e$. 
This structure called Shapiro step\cite{shapiro} is adopted to the voltage standard around the world,\cite{hamilton, kohlmann} since the voltage can be determined in the order of 10$^{-9}$ accuracy by the precise values of the frequency and the fundamental constants.\cite{hamilton, kohlmann}

On the other hand, many authors studied the Josephson effect in a hybrid structure of superconductor and ferromagnet.\cite{ryazanov_01,kontos,sellier03,robinson,robinson2,frolov}
In this {\it ferromagnetic Josephson junction} (FJJ), two superconducting electrodes are separated by a thin ferromagnetic layer, 
whose thickness is controlled in the order of nano meter.\cite{ryazanov_01,kontos,sellier03,robinson,robinson2,frolov}
If DW is put into the ferromagnetic layer of FJJ, it might be possible to measure the oscillatory DW more precisely 
by using the Josephson effect.

\begin{figure}[!t]
\begin{center}
\includegraphics[width=7.5cm]{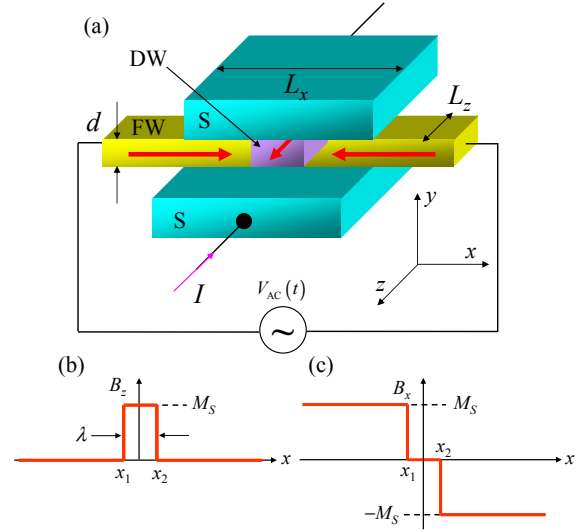}
\caption{ (Color online)
(a) Schematic illustration of ferromagnetic Josephson junction with a ferromagnetic wire (FW) with the saturation magnetization $M_{\rm S}$.  
A dc bias current $I$ drives a phase dynamics between two superconductors (S's), and  
a domain wall motion (DW) is induced by an ac voltage $V_{\rm ac}(t)$.
The following parameters of the junction, $L_{x}$=0.5 $\mu$m, $L_{z}$=200 nm, and $d$=5 nm, are used to estimate the $I$-$V$ curve.  
Red arrows are magnetizations in FW. 
(b) and (c) are the $z$- and $x$-components of magnetic flux density in FW, respectively. The DW thickness $\lambda=x_{2}-x_{1}$ is assumed to be constant as $\lambda$=100 nm. 
}
\label{fjj-gm}
\end{center}
\end{figure}
In this letter, by using FJJ, we theoretically propose a principle to measure the oscillatory DW more precisely than the present accuracy. 
As depicted in Fig.\ref{fjj-gm}(a), we use FJJ composed of a ferromagnetic Py wire with DW
and two superconducting Nb electrodes. 
The interface is parallel to the $xz$-plane with width $L_x$ and depth $L_z$. A dc bias current is applied to the $y$-direction between the electrodes separated by height $d$.
We suppose the magnetic flux densities $\bm{B}({\bm r},t)$ due to DW as shown in Figs.~\ref{fjj-gm} (b) and (c), 
where $M_{\rm S}$ is the saturation magnetization and ${\bm r} = (x,z)$. 
In such a magnetic flux density, a gauge invariant phase difference $\theta({\bm r},t)$ satisfies the differential equation,\cite{barone} 
%
$\nabla _{x,z} \theta({\bm r}, t) = -d \frac{2e}{\hbar} \bm{B}({\bm r},t) \times \bm{n}$, 
%
where $\nabla _{x,z} = (\partial /\partial _{x}, 0, \partial /\partial _{z})$ and $\bm n$ is the unit vector perpendicular to the $xz$-plane. 
The solution 
is given by, 
\begin{eqnarray}
\theta({\bm r},t) &=& \varphi(t) - \frac{2e}{\hbar} M_{\rm S} z d - \frac{e\lambda d}{\hbar}M_{\rm S}, -\frac{L_{x}}{2} < x < x_{1}
\label{theta1}, \\ 
\theta({\bm r},t) &=& \varphi(t) + \frac{2e}{\hbar} M_{\rm S} (x - \frac{x_{1} + x_{2}}{2}) d, x_{1} < x < x_{2}
\label{theta2}, \\
\theta({\bm r},t) &=& \varphi(t) + \frac{2e}{\hbar} M_{\rm S} z d + \frac{e\lambda d}{\hbar}M_{\rm S}, x_{2} < x < \frac{L_{x}}{2}
\label{theta3},
\end{eqnarray}
where $x_1$ and $x_2$ are both ends of DW. 
Below, the DW thickness $\lambda\equiv x_2-x_1$ is assumed to be constant.

The DW oscillation is driven by the external ac voltage $V_{\rm{AC}}(t)$ parallel to the $x$-direction with a certain frequency. 
Here, we assume that the width of DW is much smaller than $L_x$, and  
DW harmonically oscillates 
within the junction. 
Then, the position of $x_{1}$ is expressed to be $x_{0} \sin(\omega_{\rm DW} t)$,  
where $x_{0}$ and $\omega_{\rm DW}$ are the oscillation amplitude and frequency of DW, respectively (see Figs.~\ref{fjj-gm} (b) and (c)).  

From Eqs.~(\ref{theta1}), (\ref{theta2}), and (\ref{theta3}), the Josephson current $I_{\rm J}(t)$ is given by 
\begin{eqnarray}
I_{\rm J}(t) &=&
	j_{\rm c}
	\int_{-L_{x}/2}^{L_{x}/2}dx 
		\int_{-L_{z}/2}^{L_{z}/2}dz
			\sin \theta({\bm r},t), \nonumber \\
				&=&
				-2I_{\rm c} {\tilde x_{1}}(t) \frac{\sin(\pi \phi_{x})}{\pi \phi_{x}} \sin(\pi \phi_{z} \lambda) \cos \varphi (t) \nonumber \\
				&+&I_{\rm c} 
				\left[
				\frac{\sin(\pi \phi_{x})}{\pi \phi_{x}} \cos(\pi \phi_{z} \lambda)
				+\lambda \frac{\sin(\pi \phi_{z} \lambda)}{\pi \phi_{z} \lambda} 
				\right]
				\sin \varphi(t) \nonumber \\
				&-&I_{\rm c} \lambda \frac{\sin(\pi \phi_{x})}{\pi \phi_{x}} \sin[\varphi(t) + \pi \phi_{z} \lambda] 
\label{ij-1}, 
\end{eqnarray}
where $\phi_{i} = \Phi_{i}/\Phi_{0}$ and $\Phi_{i} = d L_{i} M_{\rm S}$ 
for $i$ = $x$, $z$. 
The Josephson critical current and its density are denoted by
$I_{\rm c}=j_{\rm c}L_{x}L_{z}$ and $j_{\rm c}$, respectively. 
The position and the width of DW are normalized as, $\tilde x_{1} = x_{1}/L_{x}$ and $\tilde \lambda = \lambda/L_{x}$.

We examine the $I$-$V$ curve in FJJ by the resistively shunted junction (RSJ) model,
\begin{equation}
I = I_{\rm J} (t) + \frac{1}{R}\frac{\Phi_{0}}{2\pi} \frac{d\varphi(t)}{dt}
\label{rsj}, 
\end{equation}
with the resistance of junction $R$ and the flux quantum $\Phi_{0}$=$h/2e$. 
\begin{figure}[!t]
\begin{center}
\includegraphics[width=6cm]{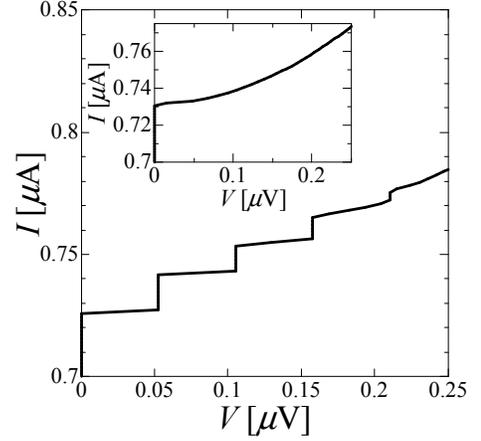}
\caption{ 
The $I$-$V$ curve is shown for $\omega_{\rm DW} = 160 {\rm MHz}$. 
The stepwise structures appear at 
$V$=$n(\hbar/2e)\omega_{\rm DW}\sim$0.052657$\times n$ $\mu$V. 
The amplitude of the DW oscillation $x_{0}/L_{x}$ is set to be 0.2. 
Without the oscillation, i.e., $\omega_{\rm DW}$=0, there is no step as shown in the inset. 
}
\label{iv-dw}
\end{center}
\end{figure}
%
%
We numerically solve Eq.$~$(\ref{rsj}) and calculate the $I$-$V$ curve 
by using the relation, $\left< \partial \varphi (t) / \partial t \right> = 2eV/\hbar $, 
where $\left< \cdot \cdot \cdot \right>$ denotes a time average. 
Figure \ref{iv-dw} shows the $I$-$V$ curve for the following parameters; $\omega_{\rm DW}$=160 MHz, $\lambda = 100$ nm, $d$=5 nm, $I_{\rm c}$=1 $\mu$A, $M_{\rm S}$=0.75 T, and $R$=1 $\Omega$.\cite{saitoh,robinson, robinson2, ando, note1}
We find the stepwise structure similar to the Shapiro step.\cite{shapiro} 
This structure, on the other hand, disappears without the DW oscillation (see the inset of Fig.~\ref{iv-dw}). 
It is noted that ferromagnetism and roughness of interface suppress the proximity effect in FJJ and then leads to the suppression of $I_{\rm c}$.\cite{ryazanov_01,kontos,sellier03,frolov,robinson,robinson2} The obtained values in Fig. 2 are experimentally observable, i.e., in order of micro-ampere and micro-volt. In view of the measurement, the fabrication of flat interface and the choice of material is important. 

To clarify the origin of the stepwise structure, we analytically examine the relation between the dc voltage and $\omega_{\rm DW}$ in the I-V curve. 
For this purpose, it is useful to note that Eq.$~$(\ref{ij-1}) is written as, 

\begin{eqnarray}
I_{\rm J}(t)  &=&
				-I_{\rm c} {\tilde x_{0}} \frac{\sin(\pi \phi_{x})}{\pi \phi_{x}} \sin(\pi \phi_{z} \lambda)
				\sum_{n=-\infty}^{\infty}
				J_{n} (\delta \varphi) \nonumber \\
				&\times&
				{\rm Im} 
				\left[
				e^{i[\varphi_{0} + ((n+1) \omega_{\rm DW} +\frac{2eV}{\hbar})t]}
				\right. \nonumber \\
				&&\left.
				-(-1)^{n} 
				e^{i[\varphi_{0} + (-(n+1) \omega_{\rm DW} +\frac{2eV}{\hbar})t]}
				\right] \nonumber \\
				&+&
				I_{\rm c} 
				\left[
				\frac{\sin(\pi \phi_{x})}{\pi \phi_{x}} \cos(\pi \phi_{z} \lambda)
				+\lambda \frac{\sin(\pi \phi_{z} \lambda)}{\pi \phi_{z} \lambda} 
				\right] \nonumber \\
				&\times& 
				\sum_{n=-\infty}^{\infty}
				J_{n} (\delta \varphi)
				{\rm Im}
				\left[
				e^{i[\varphi_{0} + (\frac{2eV}{\hbar} - n \omega_{\rm DW})t]} 
				\right] \nonumber \\
				&-&
				I_{\rm c} \lambda \frac{\sin(\pi \phi_{x})}{\pi \phi_{x}}
				\sum_{n=-\infty}^{\infty}
				J_{n} (\delta \varphi)
				{\rm Im}
				\left[
				e^{i[\varphi_{0} + \pi \varphi_{z} {\tilde \lambda} + (\frac{2eV}{\hbar} - n \omega_{\rm DW})t]}
				\right], \nonumber \\
\label{ij-2}
\end{eqnarray}
where $J_{n}(\delta \varphi)$ is the first kind Bessel function.
To obtain Eq.~(\ref{ij-2}), we suppose $\varphi(t) = \varphi_{0} + 2eVt/\hbar + \delta \varphi \sin(\omega_{\rm DW} t)$ with a constant $\varphi_{0}$. 
The time-dependent term $2eVt/\hbar$ is due to the applied bias voltage. 
The DW oscillation modulates the phase variable such as $\delta \varphi \sin(\omega_{\rm DW} t)$ with an amplitude $\delta \varphi$. 
The stepwise structure can be associated with the dc component of $I_{\rm J}(t)$, which 
is given by taking the time average of Eq.$~$(\ref{ij-2}) 
and becomes finite for $2eV/\hbar$=$\pm n\omega_{\rm DW}$. 
Therefore, the stepwise structures appear at the following voltages,  
\begin{equation}
V = n\frac{\hbar}{2e}\omega_{\rm DW}
\label{omega-j}, 
\end{equation}
with integer $n$.
It is worth to note that Eq.~(\ref{omega-j}) relates $V$ to $\omega_{\rm DW}$ only by the fundamental constant $\hbar/e$ and $n$. 
Therefore, we can precisely determine the frequency of DW oscillation by the $I$-$V$ curve.  
Those fundamental constants are known as, $e$=1.602 176 487(40)~10$^{-19}$C and $\hbar$=1.054 571 628(53)~10$^{-34}$J$\cdot $s 
with accuracy in order of $10^{-10}$.\cite{rmp80}
On the other hand, $V$ is precisely determined by the conventional Josephson junction in the order of $10^{-9}$ accuracy.\cite{hamilton, kohlmann} 
Hence, the measurement on $V$ can determine the DW frequency precisely. 
Accuracy of conventional measurement on the DW oscillation is limited in order of about $10^{-2}\sim 10^{-3}$, since those methods depend on material parameters such as anisotropic field, saturation magnetization and so on.\cite{saitoh, boone}
Therefore, our method will be more precise than the present accuracy by conventional one. 

So far, we have discussed the $I$-$V$ curve in the FJJ composed of the ferromagnetic Py wire and superconducting Nb electrodes. 
If $j_{\rm c}$ is larger than that of the Nb-based FJJ, 
one can expect a larger step height in the $I$-$V$ curve, 
since each step heights are proportional to $j_{\rm c}$ as shown in Eq.~(\ref{ij-2}).
Since $j_{\rm c}$ increases with the superconducting gap and/or the transmittance of junction, 
NbN, NbTiN and high $T_{\rm c}$ cuprates, which have larger gap compared with Nb, are possible candidates for superconducting electrodes.\cite{yang,keizer}
For instance, typical $j_{\rm c}$ of the NbTiN-based FJJ is about $1\times10^9$A/m$^2$, which is hundred times larger than that of the Nb-based FJJ.\cite{note1,keizer} 

We theoretically propose a principle for precise measurement of oscillatory domain wall (DW) by a ferromagnetic Josephson junction (FJJ), 
which is composed of a ferromagnetic wire (FW) with DW and two superconducting electrodes. 
The current-voltage curve of FJJ exhibits stepwise structures, only when DW oscillates in the FW. 
The voltage step appears at $V=n(\hbar/2e)\omega_{\rm DW}$ with the fundamental constant $\hbar/e$, integer number $n$, and the DW frequency $\omega_{\rm DW}$.  
Since the voltage is determined in the order of 10$^{-9}$ accuracy, our result provides the method to measure the oscillatory DW more precisely than the present accuracy. 
Our result Eq.~(\ref{ij-1}) can be applied to not only the oscillation but also more general dynamics of DW (i.e., the dependence of $x_{1}(t)$) 
such as uniform motion and so on. 
Those possibilities and their more detailed treatments will be examined in forthcoming work. 

The authors would like to thank J. Ieda and Hyun-Woo Lee for useful discussions and comments. 
A part of the calculation has been performed using RIKEN Integrated Cluster of Clusters (RICC). 
This work is partly supported by Grants-in-Aid for Scientific Research from MEXT 
(Grant No. 21360043, No. 22102501), 
the "K" computer project of Nanoscience Program, JST-CREST, and FIRST-Program. 



\begin{thebibliography}{}
\bibitem{note1}
In the FJJ with strong ferromagnet such as Fe, Co, or Py, 
the Josephson critical current density $j_{\rm c}$ is roughly estimated as $10^{7}$ A/${\rm m}^{2}$\cite{robinson,robinson2}. 
For instance, if the area of the junction $S_{\rm J}$ is 0.5 $\mu$m $\times$ 200 nm, 
then $I_{\rm c}$ $\approx $ 1 $\mu$A.
\end{thebibliography}
\end{document}